\documentclass{article}

\usepackage{arxiv}
\usepackage[T1]{fontenc}
\usepackage[utf8]{inputenc}
\usepackage{tikz}
\usetikzlibrary{arrows,arrows.meta}
\usepackage{xparse}
  \usepackage{amsmath}
\usepackage{amsthm}
  \usepackage{amssymb}
  \usepackage{hyperref}

  \hypersetup{
    colorlinks,
    linkcolor={blue!70!black},
    citecolor={blue!70!black},
    urlcolor={black}
}
 \usepackage{subfigure}
\usepackage{todonotes}
\usepackage{svn-multi}

\svnid{$Id: disc.tex 43301 2020-03-04 17:05:01Z ti5vt $}

\usepackage[export]{adjustbox}

\usepackage[ruled,vlined]{algorithm2e}
\SetKwFor{round}{\hspace*{-0.78ex}}{}{end}
\SetKwFunction{randomColor}{randomColor}
\SetKw{KwRet}{return}
\SetKw{Return}{return}
\SetKwInOut{Input}{input}

\graphicspath{{./figures/}}

\newcommand{\abs}[1]{{\left\lvert #1 \right\rvert}}
\newcommand{\suchthat}{\;\ifnum\currentgrouptype=16 \middle\fi|\;}

\newtheorem{theorem}{Theorem}

\newcommand{\mfont}[1]{\mathrm{#1}}
\newcommand{\outs}[0]{\mfont{OUT}}
\newcommand{\ins}[0]{\mfont{IN}}

\makeatletter
\newcommand*{\Xbar}{}%
\DeclareRobustCommand*{\Xbar}{%
  \mathpalette\@Xbar{}%
}
\newcommand*{\@Xbar}[2]{%
  \sbox0{$#1\mathrm{X}\m@th$}%
  \sbox2{$#1X\m@th$}%
  \rlap{%
    \hbox to\wd2{%
      \hfill
      $\overline{%
        \vrule width 0pt height\ht0 %
        \kern\wd0 %
      }$%
    }%
  }%
  \copy2 %
}
\makeatother

\RequirePackage{mleftright} 
\RenewDocumentCommand{\Pr}{ s o o m }{%
  \IfBooleanTF{#1}{%
    \IfNoValueTF{#2}{%
      \specrandop{Pr}[#4]%
    }{%
      \IfNoValueTF{#3}{%
        \specrandop{Pr}\mathopen{#2[}#4\mathclose{#2]}%
      }{%
        \specrandop{Pr}_{#3}\mathopen{#2[}#4\mathclose{#2]}%
      }
    }%
  }{%
    \IfNoValueTF{#2}{%
      \specrandop{Pr}\mleft[#4\mright]%
    }{%
      \specrandop{Pr}_{#2}\mleft[#4\mright]%
    }
  }%
}

\usepackage{fontawesome5}
\usepackage{listings}
\usepackage{mathtools}
\usepackage{tuhhcolor}

\lstdefinestyle{daStyle}{
  language         = C++,
  basicstyle       = \ttfamily\small,
  commentstyle     = \color{tuhh_darkgray},
  keywordstyle     = \bfseries,
  aboveskip        = 1pt,
  belowskip        = 0pt,
  fontadjust       = true,
  columns          = fullflexible,
  keepspaces       = true,
  tabsize          = 3,
  showstringspaces = false,
  mathescape       = true,
  escapechar       = €,
  morekeywords     = {concurrently,send,start,to,do,down,receive,id,while,repeat
,foreach,until,init,compute,let,wait,null,function},
  moredelim        =**[is][\color{red}]{§}{§},
  moredelim        =**[is][\sffamily\slshape]{`}{`},
  literate         = {Ä}{{\"A}}1
                     {Ö}{{\"O}}1
                     {Ü}{{\"U}}1
   	                 {ä}{{\"a}}1
			               {ö}{{\"o}}1
			               {ü}{{\"u}}1
			               {ß}{{\ss{}}}1
                     {_m}{\textsubscript{m}}2
                     {_p}{\textsubscript{p}}2
                     {_u}{\textsubscript{u}}2
                     {_v}{\textsubscript{v}}2
                     {_w}{\textsubscript{w}}2
                     {_t}{\textsubscript{t}}2
                     {_p}{\textsubscript{p}}2
                     {_id}{id}3
                     {_or}{or}3
                     {_for}{for}4
                     {_not}{not}4
                     {_and}{and}4
                     {_compute}{compute}8
                     {_receive}{receive}8
                     {_start}{start}5
                     {_wait}{wait}5
                     {_until}{until}6
                     {_and}{and}4
                     {_send}{send}5
                     {_new}{new}4
                     {_not}{not}4
                     {_to}{to}3
                     {_do}{do}3
                     {_let}{let}4
                     {_using}{using}6
                     {:}{$\;:\;$}1
                     {*}{{$\ast$}}1
                     {,}{,$\,$}1
                     {+}{{$\,+\,$}}1
                     {*}{{$\,\ast\,$}}1
                     {\%}{{$\,\%\,$}}1
                     {-}{{$\,-\,$}}1
                     {>}{{$\;>\;$}}1
                     {≥}{{$\;\ge\;$}}1
                     {>=}{{$\;\ge\;$}}2
                     {<}{{$\;<\;$}}1
                     {≤}{{$\;\le\;$}}1
                     {<=}{{$\;\le\;$}}2
                     {=}{{$\;=\;$}}1
                     {...}{{$\,\dots$}}3
                     {≠}{{$\;\neq\;$}}1
                     {/=}{{$\;\neq\;$}}2
                     {:=}{{$\;\coloneqq\;$}}2
                     {\\}{{$\,\backslash\,$}}1
                     {|}{{$\,\mid\,$}}1
                     {_NN}{{$\mathbb{N}$}}3
                     {N+(}{N\textsuperscript{+}$\!$(}3
                     {_in}{{$\!\!\in\!\!$}}2
                     {_notin}{{$\!\!\not\in\!\!$}}2
                     {__in}{\textsubscript{in}}4
                     {__out}{\textsubscript{out}}5
                     {_cup}{{$\!\cup\!$}}4
                     {_omega}{{$\omega$}}6
                     {_Delta}{{$\Delta$}}6
                     {_exists}{{$\Exists\!\!\!$}}7
                     {_forall}{{$\Forall\!\!\!$}}7
                     {_empty}{{$\emptyset$}}6
                     {_infty}{{$\infty$}}6
                     {_bot}{{$\bot$}}4
                     {_\{}{{$\{$}}2
                     {_\}}{{$\}$}}2
                     {_:}{$:$}2
                     {/\\}{{$\!\wedge\!$}}2
                     {\\/}{{$\!\vee\!$}}2
                     {<<}{{$\langle$}}2
                     {>>}{{$\rangle$}}2
                     {s.t.}{{\textbf{s.t.}}}4
                     {_unvis}{\textsubscript{unvis}}6
                     {_0}{\textsubscript{0}}1
                     {_1}{\textsubscript{1}}1
                     {_2}{\textsubscript{2}}1
                     {_3}{\textsubscript{3}}1
                     {_4}{\textsubscript{4}}1
                     {_5}{\textsubscript{5}}1
                     {_6}{\textsubscript{6}}1
                     {_7}{\textsubscript{7}}1
                     {_8}{\textsubscript{8}}1
                     {_9}{\textsubscript{9}}1
                     {_notp}{\textsubscript{$\overline{\texttt{p}}$}}5
                     {not_p}{$\overline{\texttt{p}}$}5
                     {not_parity}{$\overline{\texttt{parity}}$}5
                     {++}{{++}}2
                     {--}{{--}}2
                     {init:}{{\textbf{\textcolor{\cmdColor}{init}}:}}5
                     {start:}{{\textbf{\textcolor{\cmdColor}{start}}:}}6
                     {compute:}{{\textbf{\textcolor{\cmdColor}{compute}}:}}8
                     {Inevery:}{{\textbf{\textcolor{\cmdColor}{In every round do}}:}}8
                     {receive(}{{\textbf{\textcolor{\cmdColor}{receive}}(}}8
                     {_receive(}{{\textbf{receive}(}}9
}

\def\cmdColor{tuhh_darkred}

\newcommand{\fontmathtext}[1]{\mathsf{#1}}%
\DeclareRobustCommand{\alorithm}[1]{{\ensuremath{{\cal A}_\fontmathtext{#1}}}\xspace}
\DeclareRobustCommand{\Adeg}{\alorithm{deg}}

\title{Move Complexity of a Self-Stabilizing Algorithm for Maximal Independent Sets}
\author{%
  \href{https://orcid.org/0000-0001-9964-8816}{%
    \includegraphics[height=0.8em]{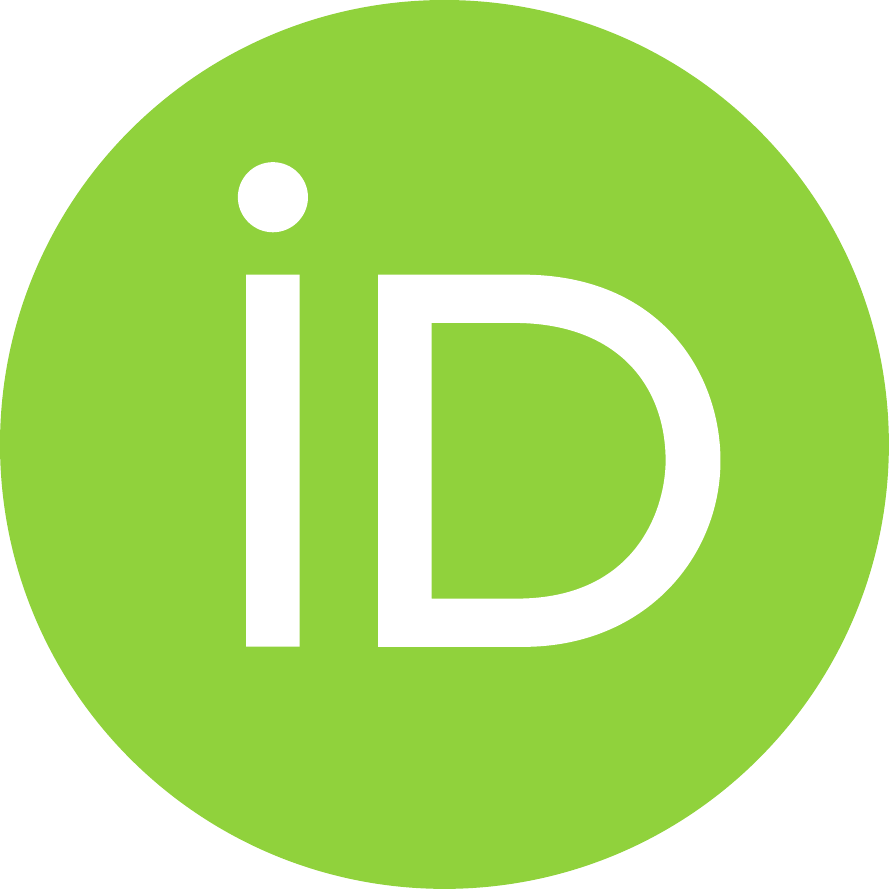}%
    \hspace{1mm}%
    Volker Turau%
  } \\
  Institute of Telematics \\
  Hamburg University of Technology \\
  21073 Hamburg, Germany \\
  \texttt{turau@tuhh.de}
}

\begin{document}

\maketitle

\begin{abstract}
  \Adeg is a self-stabilizing algorithm that computes a maximal
  independent set in a finite graph with approximation ratio
  $(\Delta + 2)/3$. In this note we show that under the central
  scheduler the number of moves of \Adeg is not bounded by a
  polynomial in $n$.
\end{abstract}


\section{Introduction}
There exist several self-stabilizing algorithms to compute a maximal
independent set in a finite graph
\cite{Ikeda:2002,Hedetniemi:2003,Turau:2007,Chiu:2014,Yen:2016}.
Algorithm \Adeg proposed by Yen et al.\ is especial because it is the
only one with a guaranteed approximation ratio. In Theorem 3.13 of
\cite{Yen:2016} the authors prove that the approximation ratio of
\Adeg is $(\Delta + 2)/3$. The authors also prove that the algorithm
stabilizes for every graph, but an upper bound for the move complexity
was not given. In this note we show that for some graphs the move
complexity grows exponentially with the size of the graph.

\section{Algorithm \Adeg}
Let $G(V,E)$ be an undirected finite graph. For $v\in V$ denote by
$N(v)$ the set of neighbors of $v$ and let $N[v]=N(v)\cup \{v\}$. The
degree $deg(v)$ of a node $v\in V$ equals $\abs{N(v)}$. Let
$\Delta = \max \{ deg(v) \suchthat v\in V\}$ and denote by
$N^{\scriptscriptstyle \le}(v)$ the set of neighbors of $v$ with
degree at most that of $v$.

Algorithm \Adeg uses a single variable $\mathit{state}$ for each node.
This variable can have the values $\ins$ and $\outs$. Let
$I^{\scriptscriptstyle \le}(v) = \{w\in N^{\scriptscriptstyle
  \le}(v)\mid w.\mathit{state} = \ins\}$. \Adeg consists of the
following two simple rules:
\begin{enumerate}
\addtolength{\itemindent}{-5pt}
\item[R1:] $\mathit{state} = \outs\,\wedge\, I^{\scriptscriptstyle \le}(v) = \emptyset ~~~~\longrightarrow state := \ins$
\item[R2:] $\mathit{state} = \ins\,\wedge\,I^{\scriptscriptstyle \le}(v) \not= \emptyset\,  ~~~~~~~\longrightarrow state := \outs$
\end{enumerate}

\Adeg stabilizes under the central scheduler and if no rule is enabled
the set $I=\{v\in V\mid v.\mathit{state} = \ins\}$ is a maximal
independent set of $G$ \cite{Yen:2016}. For regular graphs, algorithm
${\cal A}_{\mathrm{MIS}}$ coincides with the algorithm proposed in
\cite{Hedetniemi:2003}. Algorithm \Adeg preferentially places nodes
with smaller degree into state $\ins$. The intention is to find larger
independent sets. The following theorem is proved in \cite{Yen:2016}.

\begin{theorem}
  \Adeg has approximation ratio of \Adeg is $(\Delta + 2)/3$.
\end{theorem}

\section{Construction of Graphs $G_{d,w}$}
\begin{figure}[h]%
  \hfill
    \includegraphics[valign=t,scale=0.9]{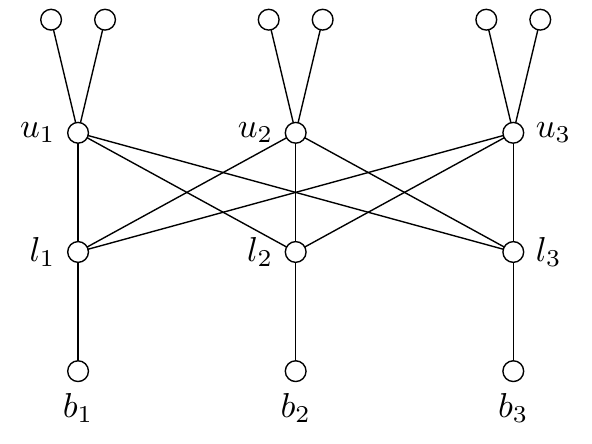}
  \hfill\null
  \caption{The  graph $G_{13}$ has $15$ nodes and $\Delta=5$.\label{fig:g13}}
\end{figure}

In this section we construct a family of graphs $G_{d,w}$ for which
\Adeg requires an exponentially growing number of moves. The
construction of $G_{d,w}$ depends on two parameters $d\ge 1$ and
$w\ge 1$. Let $H_w$ be a complete bipartite graph with $2w$ nodes. The
nodes of $H_w$ consist of the two independent sets, the upper nodes
$\{u_1,\ldots,u_w\}$ and the lower nodes $\{l_1,\ldots,l_w\}$. The
graph $G_{d,w}$ basically consists of of $d$ copies of $H_w$. These
graphs are arranged vertically and adjacent copies are connected by
$w$ edges: one edge between each node $l_i$ of one copy of $H_w$ and
node $u_i$ of the copy below. Furthermore, we attach to each node
$l_i$ of the lowest copy of $H_w$ a node $b_i$. So far we
$2wd + w=w(2d+1)$ nodes. Each node of any of the copies of $H_w$ has
degree $w+1$. Next we attach more leaves to the nodes of the subgraphs
$H_w$. The goal of this last step is to attain a graph, where the
degree of a node is one less than the degrees of the neighbors that
are one level higher. Fig.~\ref{fig:g13} shows the graph $G_{13}$.
Note that $deg(u_i) = 5$ and $deg(l_i) =
4$. 

\begin{figure}[h]%
  \hfill
    \includegraphics[valign=t,scale=0.9]{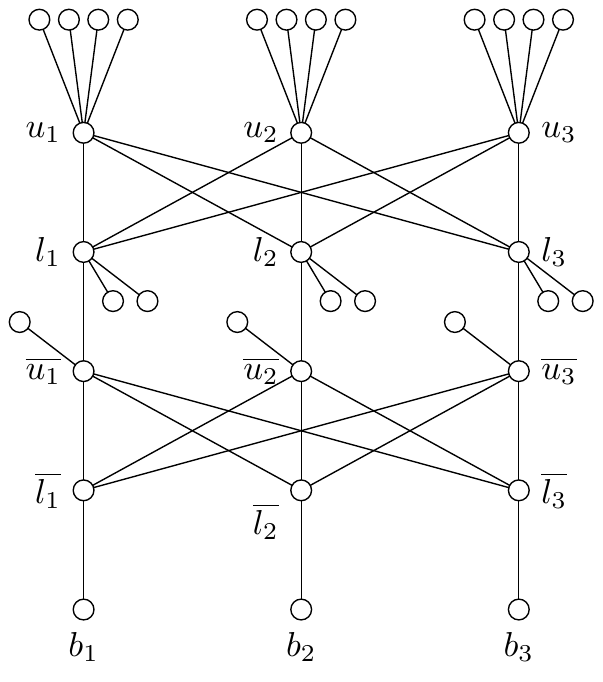}
  \hfill\null
  \caption{The  graph $G_{23}$ has $36$ nodes and $\Delta=7$.\label{fig:g23}}
\end{figure}

The number of nodes attached in the last step is equal to
\[ w\left(\sum_{i=0}^{2d-1} i + 1\right)= w(d(2d-1) +1)\] Thus, graph
$G_{d,w}$ has $w(2d+1) + w(d(2d-1)+1) = w(2d^2+d+2)$ nodes and
$\Delta=w+2d$. Fig.~\ref{fig:g23} shows the graph $G_{23}$.

\section{Executing \Adeg on $G_{dw}$}
\begin{figure*}[h]
\begin{center}
\subfigure[All $u_i$ execute rule R1.]{\includegraphics[scale=0.7]{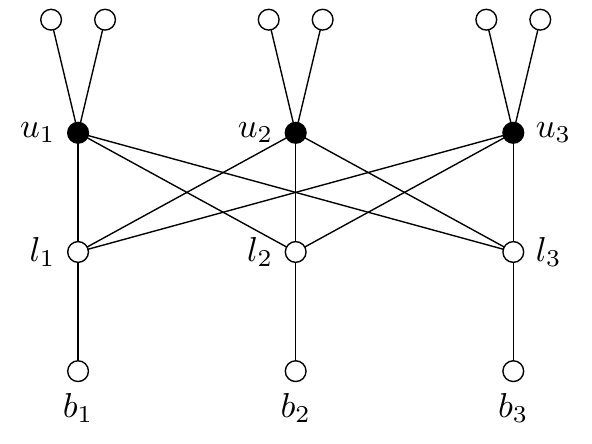}} \hspace*{10mm}
\subfigure[Node $l_1$ executes rule R1.]{\includegraphics[scale=0.7]{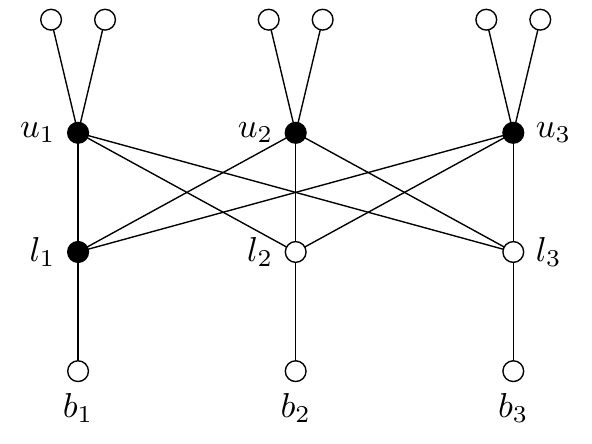}\label{ex2b}}  \hspace*{10mm}
\subfigure[All $u_i$ execute rule R2.]{\includegraphics[scale=0.7]{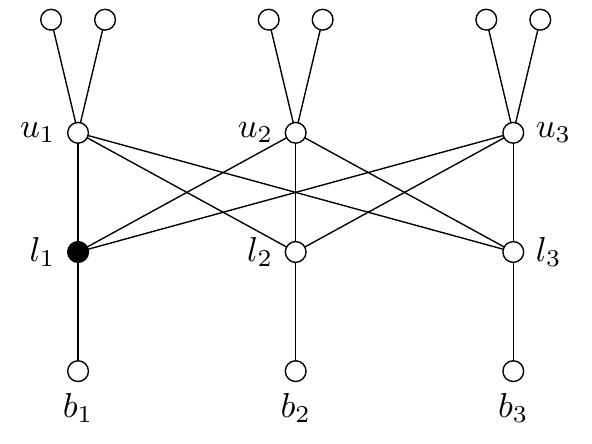}\label{ex2c}}
\\
\subfigure[Node $b_1$ executes rule R1.]{\includegraphics[scale=0.7]{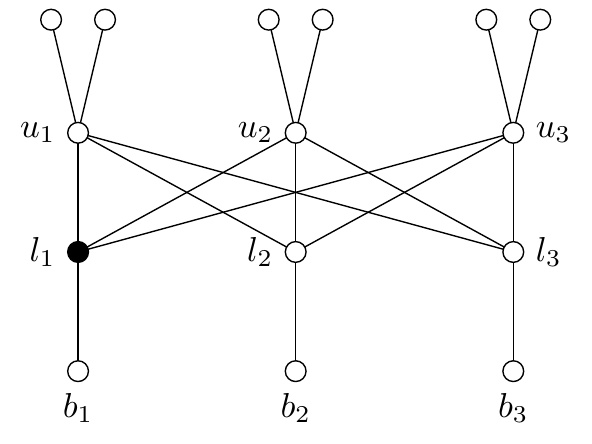}} \hspace*{10mm}
\subfigure[Node $l_1$ executes rule R2.]{\includegraphics[scale=0.7]{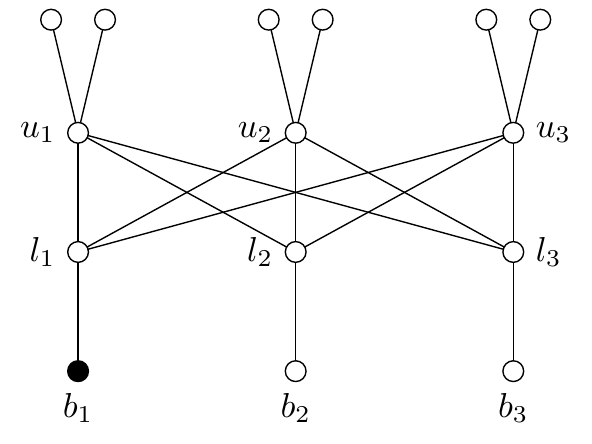}\label{ex2d}}  \hspace*{10mm}
\subfigure[All $u_i$ execute rule R1.]{\includegraphics[scale=0.7]{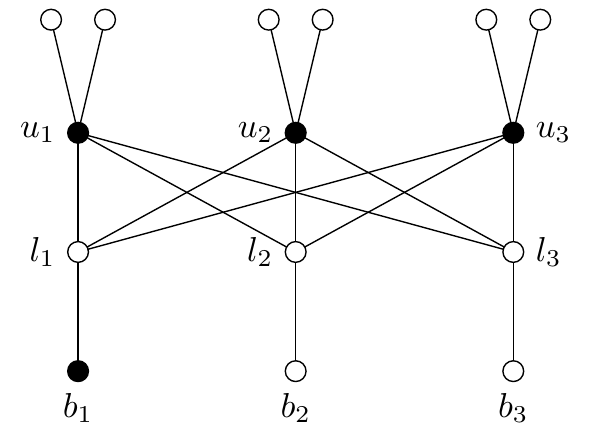}\label{ex2e}}
\caption{Execution of \Adeg requiring at least $w^2$ moves. The depicted sequence is repeated $w$ times.}\label{fig:worst_case}
\end{center}
\end{figure*}

First we consider the graph $G_{1w}$. We start with an initial
configuration where each node is in state $\outs$ (see
Fig.~\ref{fig:worst_case}). In the first $w$ steps all $u_i$ execute
rule R1 and change to state $\ins$. Then node $l_1$ also changes to
state $\ins$, note that $deg(l_1)<deg(u_i)$. This enables rule R2 for
all nodes $u_i$ and one by one changes back to state $\outs$. Then
node $b_i$ also executes rule R1 and changes to $\ins$. This forces
node $l_1$ to change back to state $\outs$. Now we are back at the
initial configuration, except for node $b_1$ which is now in state
$\ins$. Next the described process repeats itself, with $l_2$ (resp.\
$b_2$) assuming the role of $l_1$ (resp.\ $b_1$). Then we are back
again in the initial configuration except for nodes $b_1$ and $b_2$.
This sequence repeats itself $w$ times resulting in a configuration
where all nodes but $b_1,\ldots,b_w$ are in state $\outs$. In this
execution rule R1 was executed at least $w^2$ times.

Next we consider the graph $G_{2w}$ (see Fig.~\ref{fig:g23}). We start
again with an initial configuration where each node is in state
$\outs$. We can repeat the execution described for graph $G_{1w}$ for
the upper copy of the bipartite graph $H_w$ with the only difference
the nodes $u_i$ of the lower copy of $H_w$ take over the role of the
nodes $b_i$. To avoid confusion we denote the nodes of this copy of
$H_w$ by $\overline{u}_i$ and $\overline{l}_i$. Thus, we reach a
configuration where all nodes but nodes $\overline{u}_i$ are in state
$\outs$. At this point in time rule R1 has been executed at least
$w^2$ times. Then node $\overline{l_1}$ changes to state $\ins$. This
forces all nodes $\overline{u}_i$ to change back to $\outs$. Then node
$b_1$ also executes rule R1 and changes to $\ins$. This forces node
$\overline{l_1}$ to change back to state $\outs$. Now we are back at
the initial configuration, except for node $b_1$ which is now in state
$\ins$. Then the whole process beginning with the upper copy of $H_w$
repeats again. This time nodes $\overline{l_2}$ and $b_2$ take over
the role of $\overline{l_1}$ and $b_1$. Thus, so far rule R1 has been
executed at least $2w^2$ times. We can repeat the process $w$ times.
Thus, there exists an execution of \Adeg for $G_{2w}$ that contains at
least $w^3$ executions of rule R1.

This construction also works for graphs $G_{dw}$ for any value of $d$.
Hence, there is an execution of \Adeg for $G_{dw}$ that contains at
least $w^{d+1}$ moves. Let $w=d$ and $n= w(2w^2 +w+2) \in O(w^3)$.
Then $w^{d+1} = (w^3)^\frac{d+1}{3} \in n^{O(n^{1/3})}$.
Thus, there exist a graph with $O(n)$ nodes for which \Adeg
requires $n^{O(n^{1/3})}$ moves.

\begin{theorem}
Using the central scheduler the number of moves of algorithm \Adeg is
not bounded by a polynomial in $n$.
\end{theorem}

\bibliographystyle{unsrt}
\bibliography{arxiv}

\end{document}